\documentclass[conference]{IEEEtran}
\IEEEoverridecommandlockouts
\usepackage{cite}
\usepackage{amsmath,amssymb,amsfonts}
\usepackage{graphicx}
\usepackage{textcomp}
\usepackage{xcolor}
\usepackage{subfigure}
\usepackage{tablefootnote}
\usepackage{booktabs}
\usepackage{makecell}
\usepackage{stfloats}
\usepackage{pifont}
\usepackage{algorithm}  
\usepackage[normalem]{ulem}  
\usepackage{algpseudocode}  
\usepackage{amsmath}  
\usepackage{multirow}
\usepackage{fix-cm}
\usepackage{url}
\def\BibTeX{{\rm B\kern-.05em{\sc i\kern-.025em b}\kern-.08em
    T\kern-.1667em\lower.7ex\hbox{E}\kern-.125emX}}
    
\begin{document}

\title{\vspace{-0.0cm}LLM-Aided Efficient Hardware Design Automation\\
\vspace{-0.3cm}
}

\author{
\IEEEauthorblockN{Kangwei Xu\textsuperscript{1}, Ruidi Qiu\textsuperscript{1}, Zhuorui Zhao\textsuperscript{1},  Grace Li Zhang\textsuperscript{2}, Ulf Schlichtmann\textsuperscript{1}, Bing Li\textsuperscript{3}}
\IEEEauthorblockA{\textsuperscript{1}\textit{Chair of Electronic Design Automation, Technical University of Munich (TUM)}, Munich, Germany \\
\textsuperscript{2}\textit{Hardware for Artificial Intelligence Group, Technical University of Darmstadt}, Darmstadt, Germany \\
\textsuperscript{3}\textit{Research Group of Digital Integrated Systems, University of Siegen}, Siegen, Germany \\
Email: $\{$kangwei.xu, r.qiu, zhuorui.zhao, ulf.schlichtmann$\}$@tum.de, grace.zhang@tu-darmstadt.de, bing.li@uni-siegen.de}
\vspace{-1.0cm}
}

\maketitle

\begin{abstract}
%With the rapid growth in the complexity of modern chips
With the rapidly increasing complexity of modern chips, hardware engineers are required to invest more effort in tasks such as circuit design, verification, and physical implementation. These workflows often involve continuous modifications, which are labor-intensive and prone to errors. Therefore, there is an increasing need for more efficient and cost-effective Electronic Design Automation (EDA) solutions to accelerate new hardware development. Recently, large language models (LLMs) 
%have made remarkable advancements 
have made significant advancements in contextual understanding, logical reasoning, and response generation. Since hardware designs and intermediate scripts can be expressed in text format, it is reasonable to explore whether integrating LLMs into EDA could simplify and fully automate the entire workflow. Accordingly, this paper discusses such possibilities in several aspects, covering hardware description language (HDL) generation, code debugging, design verification, and physical implementation. Two case studies, along with their future outlook, are introduced to highlight the capabilities of LLMs in code repair and testbench generation. Finally, future directions and challenges are highlighted to further explore the potential of LLMs in shaping the next-generation EDA.
\end{abstract}

\section{Introduction}
\begin{figure}[]
%\vspace{-0.5cm}
\centering	\includegraphics[width=1.02\linewidth]{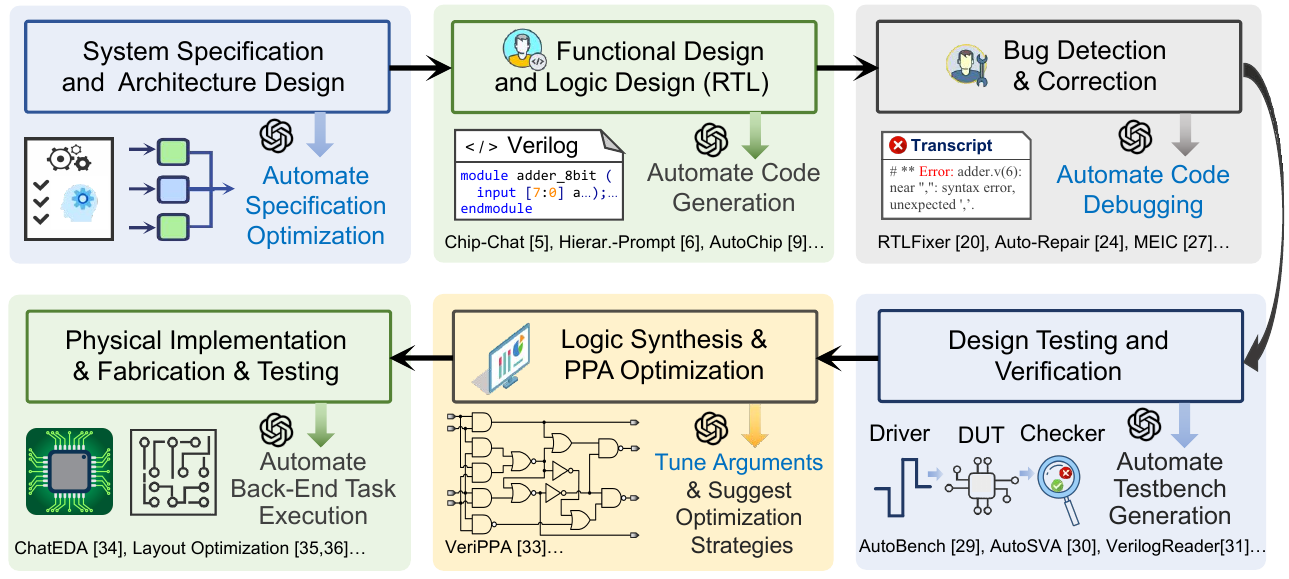}
\vspace{-0.55cm}
\caption{Typical chip design flow and potential LLM applications.}
\label{fig:eda}
\vspace{-0.4cm}
\end{figure}

Electronic Design Automation (EDA) spans the entire workflow from logic design to manufacturing, which plays a key role in improving hardware performance, reducing costs, and shortening development cycles. The advent of AI brings revolutionary changes to reshape the future of EDA \cite{b0}. By leveraging techniques such as large language models (LLMs), patterns can be learned, and insights can be inspired from a large amount of historical data, providing more efficient and intelligent solutions for the EDA workflow~\cite{b0.2, b0.3, b0.4, b1, b2, b2.1, b2.2}. Since circuits can be represented by hardware description languages (HDL), and intermediate scripts can be written in text format, integrating EDA with LLMs to release human efforts has become a promising trend. Recent studies show that the application of LLMs in hardware design demonstrates their superior capabilities and expertise in the field of EDA~\cite{b3, b4, b4.0, b4.00, b4.01, b4.02, b4.03, b4.04}.
Fig.~\ref{fig:eda} provides an overview of LLM applications in the typical chip design flow, showing their roles in automating tasks like specification optimization, RTL generation, \textit{etc.}, thereby enhancing the efficiency of the hardware design process.

While LLMs bring many benefits to the EDA workflow, they would be widely adopted only if they can achieve groundbreaking advancements to address the complexities of hardware design. The diversity of EDA tasks, along with the highly specialized nature of HDL, presents a significant challenge for applying LLMs to automate the hardware design process. In this paper, a comprehensive overview of how LLM is shaping the next-generation EDA is provided to evaluate whether this integration is an actual breakthrough or an overestimated future. 

%This paper critically examines whether the integration of EDA with LLMs heralds a new future. 

The rest of this paper is organized as follows. Section II provides a comprehensive and in-depth discussion of current trends in LLM applications in EDA. Section III analyzes two emerging cases in code repair and testbench generation. Section IV explores the future applications and challenges of LLMs in reshaping EDA, and Section V concludes the paper.

%\begin{figure}[]
%%\vspace{-0.5cm}
%\centering	\includegraphics[width=0.6\linewidth]{autochip.pdf}
%%\vspace{-0.45cm}
%	\caption{AutoChip HDL generator framework uses feedback messages from the compiler to iteratively correct the generated code~\cite{b3}.}
%	\label{fig:autochip}
%\vspace{-0.5cm}
%\end{figure}

\section{State of the Art of LLM Applications in EDA}

Hardware design typically begins with design specifications, which are then written into hardware description languages (HDL) by experienced engineers. This process is not only error-prone but also time-consuming and requires extensive human debugging efforts. Recently, LLMs have demonstrated remarkable capabilities in context understanding and logical reasoning, making it possible to automate HDL generation~\cite{b0.2, b0.3, b0.4}. 
%As shown in Fig.~\ref{fig:autochip}, 
Chip-Chat~\cite{b1} uses GPT-4 to collaborate with hardware engineers, completing a full HDL writing process for a tapeout and testing it on a new microprocessor design with an 8-bit accumulator architecture. Given the challenges of using LLMs to generate HDL for complex hardware tasks, the authors then propose hierarchical prompting techniques to enable efficient, stepwise hardware design~\cite{b2}. AutoChip~\cite{b3} introduces an automated, feedback-driven method utilizing LLMs to generate HDL. It combines LLMs with output from Verilog compilers to iteratively generate the design. An initial module is generated using design prompts. Afterward, this module is corrected based on compilation errors and simulation messages. GPT4AIGChip~\cite{b4} is a framework that automates AI accelerator design by integrating LLMs with natural language inputs. This framework decouples different functions and hardware modules of the AI accelerator, integrating LLM-friendly hardware templates and a prompt-enhanced generator to iteratively optimize the AI accelerator design. To protect hardware intellectual property (IP) from various threats,~\cite{b4.0} uses an LLM to automatically generate obfuscated code, enhancing the security of the original design. 

Another interesting application of LLMs is assisting engineers with code debugging~\cite{b4.11, b4.12, b4.13}. RTLFixer~\cite{b4.2} aims to leverage LLMs, along with Retrieval-Augmented Generation, to address syntax errors in HDL. When fixing code in a cross-disciplinary domain such as High-Level Synthesis (HLS), LLMs can more easily suffer from hallucinations and have difficulty solving unforeseen issues. \cite{b5.0} proposes an LLM-aided automatic C/C+ code repair framework for HLS, which employs a repair library and an iterative correction process to solve HLS incompatible issues while optimizing the hardware design. However, previous work has not yet successfully solved all functional errors, and there may be opportunities to utilize multiple LLMs to enhance the debugging effect of functional errors~\cite{b5.1}. MEIC~\cite{b6.1} proposes an automatic Verilog debugging framework with two LLM deployments. A debugging agent is responsible for identifying and fixing functional errors in the Register Transfer Level (RTL), and a scoring agent evaluates the quality of the fixes and scores each iteration to ensure that the code is gradually optimized.

%RTLFixer~\cite{b4.2} leverages Retrieval-Augmented Generation (RAG) paradigm to retrieve  human guidance from the database, and introduces an autonomous agent for reasoning and action planning (ReAct), enabling LLMs to act as autonomous agents in interactively debugging the Verilog code with feedback. 

For testing and verification, LLM4DV~\cite{b7} provides functional coverage point descriptions and then uses LLMs to generate input sequences, showing significant improvement in testing performance compared with random testing. Autobench~\cite{b8} utilizes LLMs to implement a hybrid test platform and self-testing system. An automated test platform evaluation framework named AutoEval is also introduced to assess the performance of the generated platforms. AutoSVA~\cite{b9} introduces an iterative framework that leverages formal verification to generate assertions from given hardware modules. VerilogReader~\cite{b10} uses LLMs as a reader to understand Verilog code and coverage, aiming to generate code coverage convergence tests. The coverage-explainer and design-under-test-explainer are also introduced to enrich prompts, thereby enhancing the LLM’s understanding of testing intent.

In the physical implementation domain, ChatEDA~\cite{b11} takes natural language as the input of LLaMA2, decomposes the process into task scheduling, script generation, and task execution, generating efficient codes for EDA tools to %perform back-end tasks, thus achieving an 
automate flow from RTL to GDSII. Moreover, an efficient LLM-driven layout design optimization is proposed~\cite{b12} to generate high-quality cluster constraints to achieve better PPA (Power, Performance and Area) of cell layout and debug the routability with the guidance of external expertise.

\begin{figure}[]
%\vspace{-0.5cm}
\centering	\includegraphics[width=1.02\linewidth]{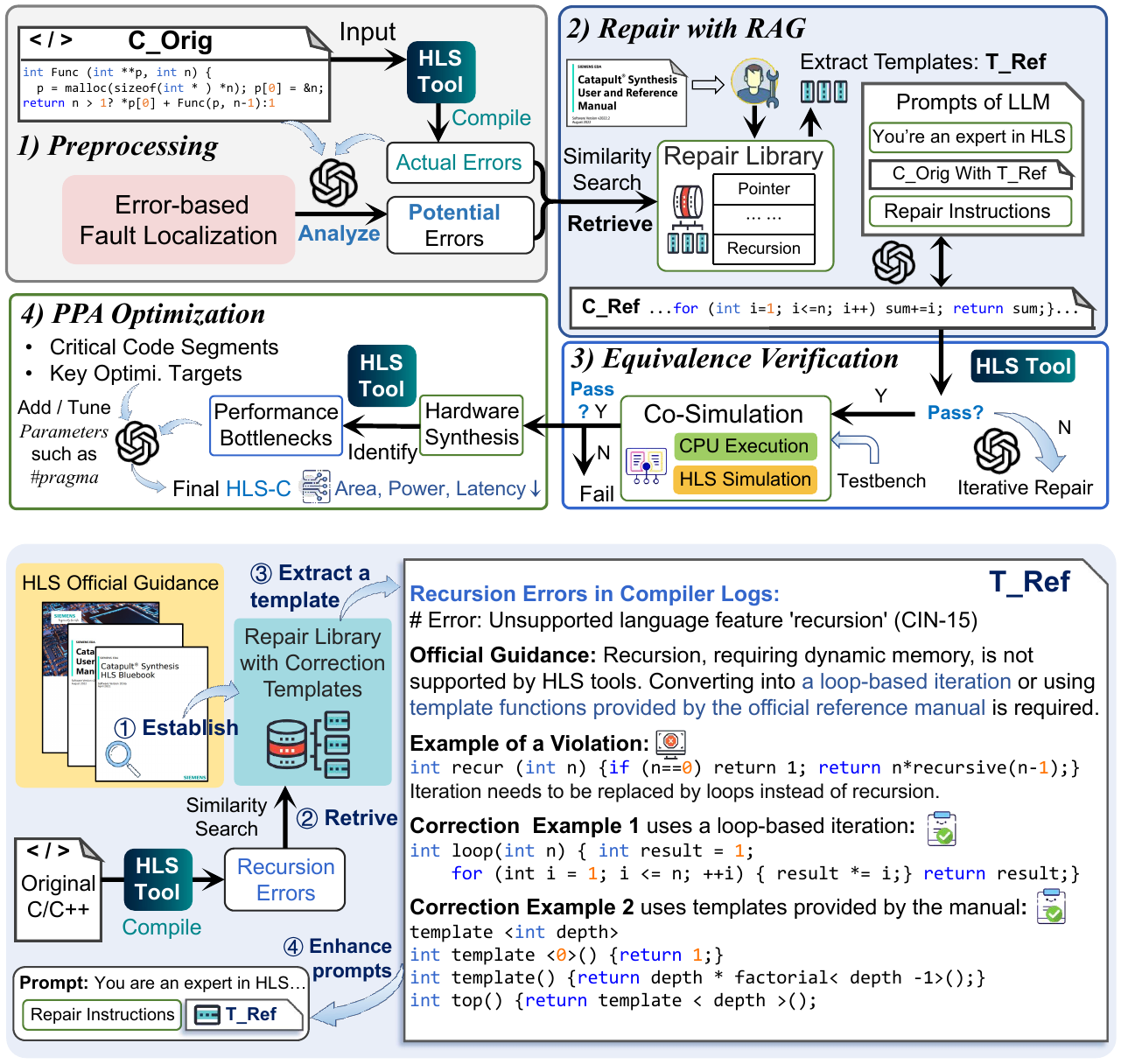}
\vspace{-0.5cm}
	\caption{HLS repair using LLM in~\cite{b5.0} with an LLM-aided code repair framework for HLS (top) and an example of using RAG to repair the recursion error (bottom). }
	\label{fig:rag}
\vspace{-0.45cm}
\end{figure}

\begin{figure*}[t]
    \centering
    \includegraphics[scale=0.955]{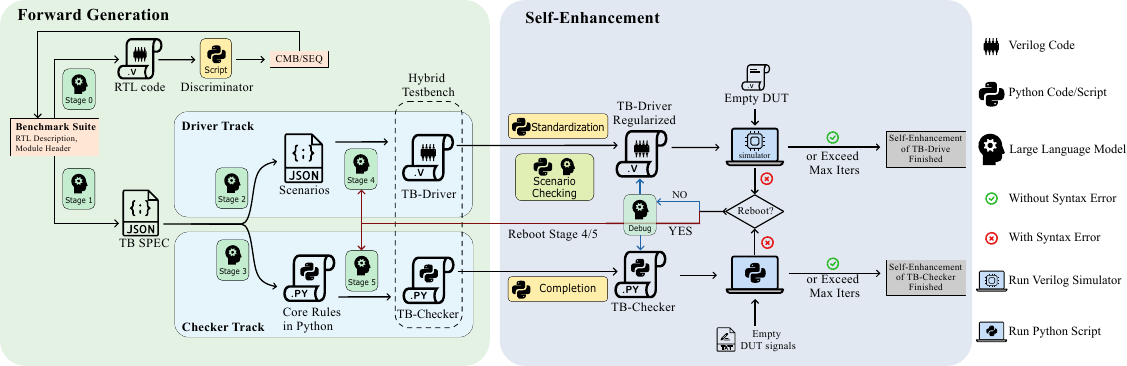}
    %\vspace{-0.25cm}
    \caption{AutoBench: Automatic testbench generation workflow for HDL design \cite{b8}.}
    \label{fig: AutoBench workflow}
    \vspace{-0.6cm}
\end{figure*}

\section{Case Study of LLMs in EDA Design Flow}

This section presents two case studies demonstrating how LLMs enhance design and testing efficiency in the EDA flow.

\subsection{LLM-Aided C/C++ Code Repair for HLS}
High-Level Synthesis (HLS) takes computer programming languages such as C/C++ as input and automatically outputs HDL like Verilog \cite{b4.5, b4.6, b5}. However, converting standard C/C++ code to HLS-compatible code (HLS-C), which can be synthesized by HLS tools, often requires tremendous manual effort. For instance, the use of dynamic memory and recursion should be rewritten by experienced engineers because hardware cannot handle unbounded data. Recently, as shown at the top of Fig.~\ref{fig:rag}, an LLM-aided code repair framework for HLS \cite{b5.0} is proposed to fix the incompatibility errors of regular C/C++ codes, which includes four stages: 

\textit{1) Preprocessing:} The original C/C++ code (C\_{Orig}), is compiled by an HLS tool, and some actual errors are returned. %Since the compiler may not be able to identify all errors in one go, an LLM is employed to detect additional potential errors.
Since the HLS compiler may not be able to detect all errors in one go, an LLM is used for other potential error detection.

\textit{2) Repair with RAG:} %Retriever-Augmented Generation (RAG) is a method that boosts the capability of the LLM by integrating external expert advice via a retriever.
Retriever-Augmented Generation (RAG) is a method of enhancing the capability of the LLM by integrating external expert knowledge via a retriever.
An example of using RAG to fix errors in HLS is illustrated at the bottom of Fig. \ref{fig:rag}, which consists of the following stages:
\ding{172} 
A repair library is created, including correction templates that each incorporate an error message, an example of the violation, contextual guidance, and correction examples extracted from the HLS manual;
\ding{173} 
After obtaining the error log from the compiler, a similarity search \cite{b5.01} is performed to match the most suitable correction template in the repair library;
\ding{174} 
Once the most similar template is retrieved, it is extracted and utilized in the prompts given to the LLM to repair the original C code (C\_Orig) and generate the repaired C code (C\_Ref);
\ding{175} 
Incorporating repair instructions and correction templates in the LLM's prompts effectively guides the LLM towards accurate repairs, thereby improving output responses.

\textit{3) Equivalence Verification:} After synthesizing the repaired C code (C\_Ref) into the corresponding RTL code, a C-RTL co-simulation is performed to verify functional equivalence.

\textit{4) PPA Optimization:} The repaired HLS-C codes are collected into a candidate list for subsequent optimization. The LLM is then employed to optimize code segments with performance bottlenecks by adjusting pragmas.

The LLM-aided C/C++ code repair framework for HLS introduced above facilitates automatic interaction between HLS tools and LLMs, thereby enhancing the efficiency of both the error correction and design optimization. Experimental results show that this framework achieves higher repair pass rates on 24 tasks compared to the direct use of the LLM. The work is open-sourced at https://github.com/code-source1/catapult.

\textit{For LLM-aided repair for HLS, 
there are still open challenges and opportunities for further exploration, which are listed as follows:} 
%\ding{172} 
a) LLM-based cross-platform agent: 
%In different HLS tools (such as the Vitis HLS tool), we hope that 
In repairing high-level codes for circuit design, knowledge and patterns of codes are usually provided in manuals of different synthesis tools. To enhance repair efficiency, it is beneficial that the HLS repair tools tool can become a cross-platform agent that automatically queries the user manuals of multiple tools and repair libraries to provide more appropriate modification suggestions and solutions.
%2) Precision-Aware Conversion: For floating-point conversions, precision loss in the software-to-hardware mapping process needs to be considered. It is worth exploring how LLMs can automatically handle bit-width conversion within a certain error margin.
%\ding{173} 
b) Hierarchical handling: It has been observed that LLMs struggle in understanding long codes. It is thus beneficial to partition the C and RTL code structures and let LLMs process submodules individually to improve accuracy and efficiency.
%4) LLM-driven Code Optimizer: LLMs could act as an optimizer, with the ability to rewrite or fine-tune code. By leveraging using appropriate optimization prompts from the user manual, LLMs could iterate multiple times, quickly determining the most effective optimization strategies.
%\ding{174} 
c) Automated testbench generation for HLS repair: In C-RTL co-simulation, testbenches are typically generated manually. 
Future research is needed to generate special testbenches, probably on-the-fly according to the LLM's understanding of the target codes, to verify individual repairs in the automated flow. These testbenches are orthogonal with the traditional function-based testbenches and can specifically target to match the patterns in the high-level programming languages, the repaired c codes, and the resulting RTL codes to correct logic errors and RTL-specific timing issues, thus leading to correct and efficient RTL designs.

\newcommand{\github}{\url{https://github.com/AutoBench/AutoBench}}

\subsection{LLM-Aided TestBench Generation for HDL Design}

Simulation-based functional verification using a testbench is a crucial phase in the design of digital hardware. 
A testbench consists of two primary parts: the \textit{driver} and the \textit{checker}. The driver is responsible for generating test stimuli and directing the Design Under Test (DUT) to produce outputs. Subsequently, the checker captures these signals from the DUT and verifies the DUT's correctness. Traditional methods for testbench generation primarily automate the design of drivers with random test stimuli, while checkers remain manually designed due to their task-specific nature. Moreover, the use of randomly generated test stimuli can be inefficient during debugging since they lack additional context information. 

Motivated by the significant performance of LLMs in circuit design, the first automatic and systematic testbench generation framework, AutoBench, was introduced in \cite{b8} to automate the design processes of both drivers and checkers, as illustrated in Fig.~\ref{fig: AutoBench workflow}. 
The framework's only input is the RTL specification (SPEC) in natural language. Ultimately, it produces a hybrid testbench comprising both Python and Verilog codes. The workflow begins with identifying the circuit type from the given RTL SPEC, based on an imperfect RTL code generated by LLM using the RTL SPEC, as outlined in \textbf{Stage 0} of Fig.~\ref{fig: AutoBench workflow}. Subsequently, in \textbf{Stage 1}, the RTL SPEC is translated into a TB SPEC by the LLM to support the subsequent stages.

\subsubsection{Design of the driver}
The driver code is responsible for generating test stimuli and directing the DUT. Here, the driver code comprises several test scenarios, each containing one or more test stimuli. In \textbf{Stage 2}, a list of test scenarios in natural language is produced by the LLM using the TB SPEC to facilitate comprehension in the subsequent stage. In \textbf{Stage 4}, the LLM is provided with both the SPEC and the scenario list to generate the complete testbench driver code in Verilog.

\subsubsection{Design of the checker}
The checker code in AutoBench is implemented in Python rather than Verilog due to Python's higher level of abstraction. Additionally, LLMs, leveraging a more extensive training dataset, exhibit superior performance in Python programming. The LLM generates the Python driver in two steps: it first creates the core functions in \textbf{Stage 3} and then produces the complete code in \textbf{Stage 5}.

\subsubsection{Self-Enhancement}
With the generated driver and checker, a self-enhancement process is undertaken to improve the testbench's correctness, as depicted in the right part of Fig.~\ref{fig: AutoBench workflow}. Automatic code standardization and completion are performed according to predetermined formatting rules. For the checker, scenario checking is conducted to ensure that all scenarios in the list are present in the final driver code. Subsequently, both codes are executed to detect any syntax errors. The LLM attempts to resolve these errors using debugging information. If it fails after several attempts, the system reboots at the code generation process from Stage 4 or Stage 5, depending on the code type.

The final revised and corrected codes will constitute the hybrid testbench produced by AutoBench. The implementation of AutoBench is accessible as open-source on \github{}. Future work will involve exploring functional and coverage-aware self-correction and self-correction mechanisms to further enhance the accuracy of the generated testbenches.

\section{Future Directions and challenges}
The application of LLMs in EDA has made significant progress. However, due to the complexity of EDA tasks, this field is still in its early stage, and the integration of LLMs into EDA faces many challenges and thus opportunities. %Modern hardware design flows mainly consist of several key stages, such as system specification, logic design, design testing, physical implementation, and so on. Each stage is essential for ensuring the successful mass production of chips. 

\begin{figure}[]
%\vspace{-0.5cm}
\centering	\includegraphics[width=1.01\linewidth]{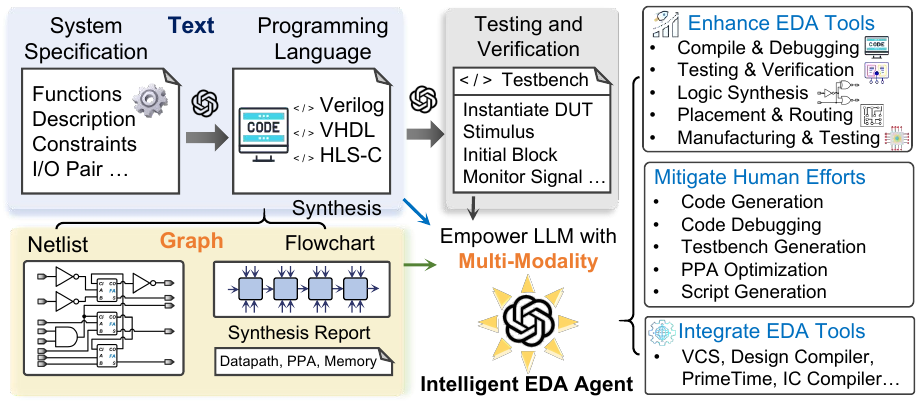}
\vspace{-0.5cm}
	\caption{LLM-Powered intelligent EDA agent.}
	\label{fig:agent}
\vspace{-0.45cm}
\end{figure}

As shown in Fig.~\ref{fig:agent}, we envision an LLM-powered intelligent agent that will be built for EDA, which can integrate various stages of the EDA workflow and provide general solutions throughout the entire process. This agent could be developed by fine-tuning open-source LLMs or leveraging the commercial LLMs to enhance the performance of EDA tools. Not only specifications and HDL could be embedded in LLMs, but multi-modal visual information (schematics, flowcharts, etc.) could also be used to enrich the hardware information to help LLMs better understand design intent. The agent would also seamlessly integrate across various tools, enabling full automation of the EDA workflow and reducing human efforts. Here, we propose several directions that can be further explored to advance the development of LLMs in EDA.

$\circ$~\textit{\textbf{Automated Specification Optimization:}} Writing clear and accurate architecture specifications is a critical first step in chip design, usually implemented by experienced engineers. However, any ambiguities or mistakes in the specifications may cause significant issues in subsequent design stages. Automating specification optimization presents a new opportunity 
%to improve the accuracy and efficiency of the entire workflow.
to enhance the efficiency and accuracy of the whole workflow.

$\circ$~\textit{\textbf{Multi-Modal Analysis:}} Modern hardware design contains multiple levels of details, from the logic level to the architecture level. A single modality (such as text or code) may not provide a comprehensive understanding of the design. 
%Multi-modal models can process different types of data simultaneously
Multi-modal models have the ability to process various types of data simultaneously, such as visual information (schematics, flowcharts, etc.), bridging the gap left by single-modality. Text information provides hardware logic descriptions, while visual information helps identify signal paths, dataflow, and structure, which can help LLMs better understand hierarchical circuit designs.

$\circ$~\textit{\textbf{Automated Debugging for HLS:}} In the C-RTL co-simulation of HLS, engineers often manually create test vectors and write testbenchs to identify errors in the RTL designs. Once C code is synthesized into RTL designs, timing issues may occur, which are not covered in the higher-level C/C++ code. By combining LLM capabilities with traditional HLS debugging processes, real-time testing support could be realized to correct the RTL design. In addition, accurate assertions could be generated by LLMs and integrated into formal verification workflows to enhance hardware verification.

$\circ$~\textit{\textbf{Logic Synthesis:}} Converting HDL (such as Verilog) into optimized gate-level representations (such as netlist) requires extensive tuning of the flow, including selecting optimization strategies and adjusting arguments. Achieving efficient design space exploration is challenging due to the exponential growth in the number of possible optimization combinations. By taking HDL along with high-quality prompts as input, LLMs could generate optimization flow and corresponding arguments. Moreover, the reported PPA can be fed back to the LLM, enabling improvement of the hardware design.

$\circ$~\textit{\textbf{Privacy and Security:}} Engineers may rely on cloud-based LLMs to automatically generate hardware solutions, which could expose sensitive data if transmitted without encryption. This could lead to privacy breaches or intellectual property theft. Moreover, malicious code or hardware trojans may be inserted into the generated hardware designs on the cloud platform. Deploying LLMs on local servers, combined with encrypted communication channels for interaction with cloud-based models, would protect sensitive data while allowing flexible use of cloud resources for complex tasks.

$\circ$~\textit{\textbf{Seamless Integration of EDA Tools:}} Compatibility between existing tools remains a critical issue. Differences in interfaces and data formats across various tools pose significant challenges to EDA tool integration. Therefore, an intelligent EDA system is needed to autonomously manage workflows across all stages. Real-time feedback and comprehensive results can also be fed back to the system to iteratively optimize the hardware design.

%These future directions highlight the potential for LLMs to revolutionize the EDA process. Further research in these areas will help unlock the full potential of LLMs in EDA.

\section{Conclusion}
%Integrating LLMs into EDA workflow opens up opportunities to explore a wider range of 
Integrating LLMs into EDA workflow provides opportunities for exploring a wider range of automatic design processes such as HDL generation, code debugging and testbench generation, which not only reduces hardware development costs but also shortens the time to market. Despite challenges such as hallucinations of LLMs, complexity of hardware design and data privacy concerns, etc., the application of LLMs in EDA still presents a groundbreaking opportunity, potentially paving a way for more efficient and intelligent hardware development in reshaping the future of EDA.

\vspace{12pt}


\begin{thebibliography}{00}
\bibitem{b0} Tinghuan Chen, Grace Li Zhang, Bei Yu, Bing Li, Ulf Schlichtmann, “Machine Learning in Advanced IC Design: A Methodological Survey,” IEEE Design \& Test, 2023.
\bibitem{b0.2} Kaiyan Chang, Kun Wang, Nan Yang, Ying Wang, Yinhe Han, Huawei Li, Xiaowei Li \textit{et al.}, “Data is all you need: Finetuning LLMs for Chip Design via an Automated design-data augmentation framework,” IEEE/ACM Design Automation Conference (DAC), 2024.
\bibitem{b0.3} Prashanth Vijayaraghavan, Ehsan Degan \textit{et al.}, “Chain-of-Descriptions: Improving Code LLMs for VHDL Code Generation and Summarization,” IEEE/ACM International Symposium on Machine Learning for Computer-Aided Design (MLCAD), 2024.
\bibitem{b0.4} Christopher Batten, Nathaniel Pinckney, Mingjie Liu, Haoxing Ren, Brucek Khailany, “PyHDL-Eval: An LLM Evaluation Framework for Hardware Design Using Python-Embedded DSLs,” IEEE/ACM International Symposium on Machine Learning for Computer-Aided Design (MLCAD), 2024.
\bibitem{b1} Jason Blocklove, Siddharth Garg, Ramesh Karri, Hammond Pearce, “Chip-Chat: Challenges and Opportunities in Conversational Hardware Design,” IEEE/ACM Workshop on Machine Learning for Computer-Aided Design (MLCAD), 2023.
\bibitem{b2} Andre Nakkab, Sai Qian Zhang, Ramesh Karri, Siddharth Garg, “Rome was Not Built in a Single Step: Hierarchical Prompting for LLM-based Chip Design,” IEEE/ACM International Symposium on Machine Learning for Computer-Aided Design (MLCAD), 2024.
\bibitem{b2.1} Shailja Thakur, Baleegh Ahmad, Zhenxing Fan, Hammond Pearce, Benjamin Tan, Ramesh Karri, Brendan Dolan-Gavitt, Siddharth Garg, “Benchmarking Large Language Models for Automated Verilog RTL Code Generation," IEEE/ACM Design, Automation \& Test in Europe Conference \& Exhibition (DATE), 2023.
\bibitem{b2.2} Mingjie Liu, Nathaniel Pinckney, Brucek Khailany, Haoxing Ren, “VerilogEval: Evaluating large language models for verilog code generation,” IEEE/ACM International Conference on Computer Aided Design (ICCAD), 2023.
\bibitem{b3} Shailja Thakur, Jason Blocklove, Hammond Pearce, Benjamin Tan, Siddharth Garg, Ramesh Karri, “AutoChip: Automating HDL Generation Using LLM Feedback,” arXiv preprint: 2311.04887, 2023.
\bibitem{b4} Yonggan Fu, Yongan Zhang, Zhongzhi Yu, Sixu Li, Zhifan Ye, Chaojian Li, Cheng Wan, Yingyan Lin, “GPT4AIGChip: Towards Next-Generation AI Accelerator Design Automation via Large Language Models,” IEEE/ACM International Conference on Computer Aided Design (ICCAD), 2023.
\bibitem{b4.0} Banafsheh Saber Latibari, Soheil Salehi \textit{et al.}, “Automated Hardware Logic Obfuscation Framework Using GPT," IEEE Dallas Circuits and Systems Conference (DCAS), 2024.
\bibitem{b4.00} Yu-Zheng Lin, Soheil Salehi \textit{et al.}, “HW-V2W-Map: Hardware Vulnerability to Weakness Mapping Framework for Root Cause Analysis with GPT-assisted Mitigation Suggestion,” arXiv preprint: 2312.13530, 2023.
\bibitem{b4.01} Jason Blocklove, Siddharth Garg, Ramesh Karri, Hammond Pearce, “Evaluating LLMs for Hardware Design and Test,” arXiv preprint: 2405.02326, 2024.
\bibitem{b4.02} Wenhao Sun, Bing Li, Grace Li Zhang, Xunzhao Yin, Cheng Zhuo, Ulf Schlichtmann, “Classification-Based Automatic HDL Code Generation Using LLMs,” arXiv preprint: 2407.18326, 2024.
\bibitem{b4.03} Kangwei Xu, Grace Li Zhang, Ulf Schlichtmann, Bing Li, “Logic Design of Neural Networks for High-Throughput and Low-Power Applications,” IEEE/ACM Asia and South Pacific Design Automation Conference (ASP-DAC), 2024.
\bibitem{b4.04} Yunwei Mao, You You, Xiaosi Tan, Yongming Huang, Xiaohu You, Chuan Zhang, “FLAG: Formula-LLM-Based Auto-Generator for Baseband Hardware," IEEE International Symposium on Circuits and Systems (ISCAS), Singapore, Singapore, 2024
\bibitem{b4.11} Lily Jiaxin Wan, Yingbing Huang, Yuhong Li, Hanchen Ye, Jinghua Wang, Xiaofan Zhang, Deming Chen, “Software/Hardware Co-design for LLM and Its Application for Design Verification," IEEE/ACM Asia and South Pacific Design Automation Conference (ASP-DAC), 2024.
\bibitem{b4.12} Mingjie Liu, Yun-Da Tsai, Wenfei Zhou, Haoxing Ren, “CraftRTL: High-quality Synthetic Data Generation for Verilog Code Models with Correct-by-Construction Non-Textual Representations and Targeted Code Repair,” arXiv preprint: 2409.12993, 2024.
\bibitem{b4.13} Luca Collini, Siddharth Garg, Ramesh Karri, “C2HLSC: Can LLMs Bridge the Software-to-Hardware Design Gap?,” IEEE International Workshop on LLM-Aided Design (LAD), 2024.
\bibitem{b4.2} Yun-Da Tsai, Mingjie Liu, Haoxing Ren, “Automatically Fixing RTL Syntax Errors with Large Language Model,” IEEE/ACM Design Automation Conference (DAC), 2024.
\bibitem{b4.5} Yunsheng Bai, Atefeh Sohrabizadeh, Zongyue Qin, Ziniu Hu, Yizhou Sun, Jason Cong, “Towards a Comprehensive Benchmark for High-Level Synthesis Targeted to FPGAs,” Advances in Neural Information Processing Systems (NeurIPS), 2023.
\bibitem{b4.6} Atefeh Sohrabizadeh, Cody Hao Yu, Min Gao, and Jason Cong, “AutoDSE: Enabling Software Programmers to Design Efficient FPGA Accelerators,” ACM Transactions on Design Automation of Electronic Systems (TODAES), 2022.
\bibitem{b5} Stefan Abi-Karam, Rishov Sarkar, Allison Seigler, Sean Lowe, Zhigang Wei, Hanqiu Chen, Nanditha Rao, Lizy John, Aman Arora, Cong Hao, “HLSFactory: A Framework Empowering High-Level Synthesis Datasets for Machine Learning and Beyond,” IEEE/ACM International Symposium on Machine Learning for Computer-Aided Design (MLCAD), 2024.
\bibitem{b5.0} Kangwei Xu, Grace Li Zhang, Xunzhao Yin, Cheng Zhuo, Ulf Schlichtmann, and Bing Li, “Automated C/C++ Program Repair for High-Level Synthesis via Large Language Models,” IEEE/ACM International Symposium on Machine Learning for Computer-Aided Design (MLCAD), 2024.
\bibitem{b5.01} Nils Reimers, Iryna Gurevych, “Sentence-BERT: Sentence Embeddings using Siamese BERT-Networks,” ACL Empirical Methods in Natural Language Processing (EMNLP), 2019.
\bibitem{b5.1} Wenji Fang, Mengming Li, Min Li, Zhiyuan Yan, Shang Liu, Hongce Zhang, Zhiyao Xie, “AssertLLM: Generating and Evaluating Hardware Verification Assertions from Design Specifications via Multi-LLMs,” arXiv preprint: 2402.00386, 2024.

\bibitem{b6.1} Ke Xu, Jialin Sun, Yuchen Hu, Xinwei Fang, Weiwei Shan, Xi Wang, Zhenling Jiang, “MEIC: Re-thinking RTL Debug Automation using LLMs,” IEEE/ACM International Conference on Computer Aided Design (ICCAD), 2024.
\bibitem{b7} Zixi Zhang, Greg Chadwick, Hugo McNally, Yiren Zhao, Robert Mullins, “LLM4DV: Using Large Language Models for Hardware Test Stimuli Generation,” Advances in Neural Information Processing Systems (NeurIPS) Workshop, 2022.
\bibitem{b8} Ruidi Qiu, Grace Li Zhang, Rolf Drechsler, Ulf Schlichtmann, Bing Li, “AutoBench: Automatic Testbench Generation and Evaluation Using LLMs for HDL Design,” IEEE/ACM International Symposium on Machine Learning for Computer-Aided Design (MLCAD), 2024.
\bibitem{b9} Marcelo Orenes-Vera, Margaret Martonosi, David Wentzlaff, “Using LLMs to Facilitate Formal Verification of RTL,” arXiv preprint: 2309.09437, 2023.
\bibitem{b10} Ruiyang Ma, Yuxin Yang, Ziqian Liu, Jiaxi Zhang, Min Li, Junhua Huang, Guojie Luo, “VerilogReader: LLM-Aided Hardware Test Generation,” IEEE International Workshop on LLM-Aided Design (LAD), 2024.
\bibitem{b10.0} Chuangtao Chen, Grace Li Zhang, Xunzhao Yin, Cheng Zhuo, Ulf Schlichtmann, Bing Li, “LiveMind: Low-latency Large Language Models with Simultaneous Inference,” arXiv preprint: 2406.14319, 2024.
\bibitem{b10.1} Kiran Thorat, Jiahui Zhao, Yaotian Liu, Hongwu Peng, Xi Xie, Bin Lei, Jeff Zhang, Caiwen Ding, “Advanced Large Language Model (LLM)-Driven Verilog Development: Enhancing Power, Performance, and Area Optimization in Code Synthesis,” arXiv preprint: 2312.01022, 2023.

\bibitem{b11} Zhuolun He, Haoyuan Wu, Xinyun Zhang, Xufeng Yao, Su Zheng, Haisheng Zheng, Bei Yu, “ChatEDA: A Large Language Model Powered Autonomous Agent for EDA," IEEE Transactions on Computer-Aided Design of Integrated Circuits and Systems, 2024.
\bibitem{b12} Chia-Tung Ho, Haoxing Ren, “Large Language Model (LLM) for Standard Cell Layout Design Optimization,” IEEE International Workshop on LLM-Aided Design (LAD), 2024.
\bibitem{b13} Bingyang Liu, Haoyi Zhang, Xiaohan Gao, Zichen Kong, Xiyuan Tang, Yibo Lin, Runsheng Wang, Ru Huang, “LayoutCopilot: An LLM-powered Multi-agent Collaborative Framework for Interactive Analog Layout Design,” arXiv preprint: 2406.18873, 2024.
\end{thebibliography}
\end{document}